\documentclass[prl,reprint,citeautoscript]{revtex4-2}
\usepackage[utf8]{inputenc}
\usepackage{graphicx}% Include figure files
\usepackage{dcolumn}% Align table columns on decimal point
\usepackage{framed}
\usepackage[compact,tiny,center]{titlesec}
\usepackage{xcolor}
\usepackage{fancyvrb}

\newcommand\red[1]{\color{red}#1}

\begin{document}
\title{Natural Language Processing Models That Automate Programming Will Transform Chemistry Research and Teaching}
\author{Glen M. Hocky}
\affiliation{Department of Chemistry, New York University}
\email{hockyg@nyu.edu}
\author{Andrew D. White}
\affiliation{Department of Chemical Engineering, University of Rochester}
\email{andrew.white@rochester.edu}
\date{\today}

\noindent
\begin{abstract}
Natural language processing models have emerged that can generate usable software and automate a number of programming tasks with high fidelity. 
These tools have yet to have an impact on the chemistry community. Yet, our initial testing demonstrates that this form of Artificial Intelligence is poised to transform chemistry and chemical engineering research. Here, we review developments that brought us to this point, examine applications in chemistry, and give our perspective on how this may fundamentally alter research and teaching.
\end{abstract}

\maketitle

\begin{figure}[ht]
    \centering
    \includegraphics[width=3in]{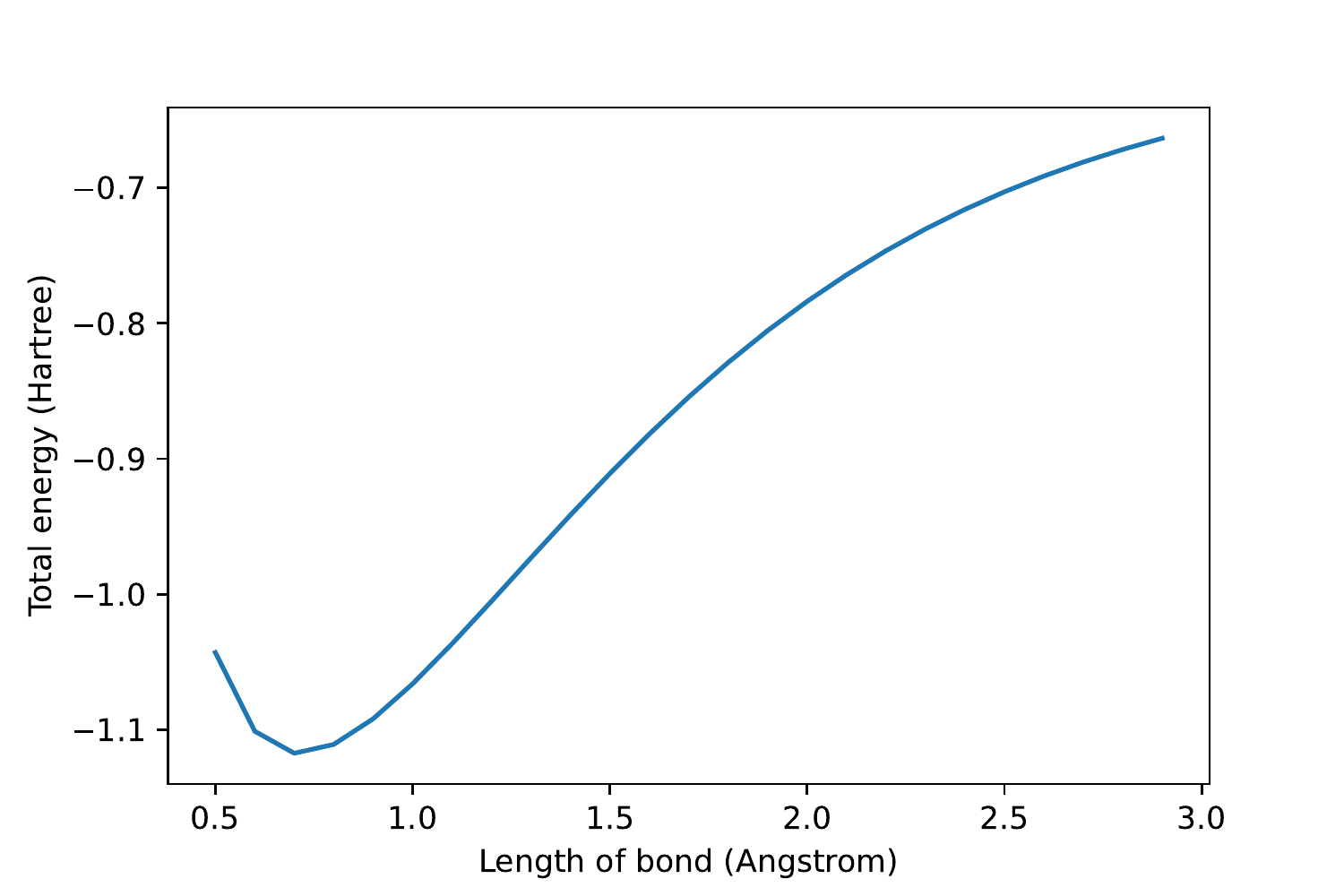}
    \caption{Prompt: \texttt{Compute the dissociation curve of H2 using the pyscf library}. See SI for code. Note, when repeating this prompt, the method, labels, and range can change due to under-specification of the request.}
    \label{fig:h2curve}
\end{figure}

\noindent
In 2021, Chen et al. released a new natural language processing (NLP) model called Codex that can generate code from natural language prompts \cite{chen2021evaluating}. Interest has been broadly focused on its application to software engineering. We, somewhat sarcastically, asked it to ``Compute the dissociation curve of H2 using pyscf'' \cite{sun2018pyscf} and the result is shown in Fig.~\ref{fig:h2curve}. It generated correct code and even plotted it (see SI for further details). Some may scoff at the artificial intelligence (AI) selected method (Hartree--Fock) and basis set (STO-3G). Thus, we asked it to ``use the most accurate method'' as a continuation of our ``conversation" and it switched to CCSD in a large basis.
AI models that can connect natural language to programming will have significant consequences for the field of chemistry---here we outline a brief history of these models and our perspective on where these models will take us.

\section{Recent developments}
There has been a flurry of advances in the topic of ``autocomplete'' style language models that can generate text given a prompt over the last three years.
These language models are deep neural networks with a specific architecture called transformers \cite{vaswani2017attention,devlin2018bert}.
These models are trained on text that has words hidden \footnote{more generally, ``tokens'' are masked}, and have the task of filling in missing text \cite{devlin2018bert,taylor1953cloze,dai2015semi}. This is called ``pre-training,'' because these models were not intended to fill in missing words, but rather be used on downstream tasks like classifying sentiment in text or categorizing text \cite{devlin2018bert}. 
Surprisingly, it was found that these models could generate a long seemingly real passage of text simply from a short initial fragment of text called a prompt \cite{devlin2018bert,radford2019language}.
These prompts can be to answer a question, summarize a story, or make an analogy---all with the same model. This was interesting, especially because the quality was beyond previous text generation methods like recurrent neural networks or hidden Markov models \cite{sutskever2011generating}.
After increasing model size and the training corpus, the next generation of language models were able to answer novel prompts beyond standard question-and-answer or writing summaries \cite{brown2020language}.
For example, given three worked out examples of extracting compound names from a sentence, the GPT-3 model could do the same for any new sentence. We show the utility of this for parsing chemistry literature in Fig.~\ref{box:chem} using text from Ref.~\citenum{hueckel2020ionic} (see SI for full details).
This result is remarkable because it requires no additional training, just the input prompt -- literally a training size of 3. Not so long ago, this was considered a difficult problem even when using thousands of training examples \cite{krallinger2015chemdner}. A caveat to these large language models (LLMs) is that they have a limited \textit{understanding} of the text which they parse or generate; for example, we find they can generate seemingly valid chemistry text but cannot answer simple questions about well known chemical trends.  

After these new LLMs were developed, anyone could have state-of-the art performance on language tasks simply by constructing a few examples of their task. In the last few months, even the need for worked out examples can be removed. In some cases a simple `imperative' sentence is enough \footnote{unpublished, but part of ongoing work known as \texttt{davinci-instruct} GPT-3 variant}. For example, a variation on the name of this article  was generated by asking an imperative-style model to ``write an exciting title" given an earlier version of the abstract.
The pace has been nothing short of remarkable, going from the transformer in 2017 to a near universal language model in 2020 to a model which can take instructions in 2021.

\begin{figure}
    \begin{framed}
    \raggedright
    \texttt{Sentence: In brief, fluorescent dyes dissolved in tetrahydrofuran (THF) were added to particles stabilized with Pluronic F108 (MilliporeSigma) to a final concentration of 30\% v/v THF, then diluted by a factor of five before washing the particles via multiple sedimentation and resuspension cycles to set them in pure water.\vspace{0.1cm}
Chemical Entities:} \textbf{{\color{red} Pluronic F108, tetrahydrofuran, THF}}
    \end{framed}
    \caption{Example of chemical entity recognition after training on three examples with GPT-3. Prompt including a direct quote of text from Ref.~\citenum{hueckel2020ionic} is in monospace and response is bolded and red. Note that this misses the connection between tetrahydrofuran and THF, and does not associate water with a chemical entity.}
    \label{box:chem}
\end{figure}

The largest and arguably most accurate model in this class is still the GPT-3 model from OpenAI \cite{brown2020language}.
GPT-3 is an enigma in the field of natural language models.
It is democratizing because anyone can create a powerful language model in a few hundred characters that is deployable immediately.
Yet its weights are a pseudo-trade secret, owned and licensed by OpenAI exclusively to Microsoft.
Thus the only way to run it is via their website (or API). These kinds of models are known as Large Language Models.
Any state-of-the-art language models should start with a LLM like GPT-3 or, for example, the freely available GPT-NEO \cite{gptneo}. 
GPT-3 has been trained on billions of tokens and no effort has yet to match its scale of training data and model size.
It can be unsettling too because it has quite adeptly captured the racism, sexism, and bias in human writing and can be reflected in its responses \cite{bender2021dangers}. Mitigating this is an ongoing effort \cite{solaiman2021process}. 
Another interesting outcome is that ``prompt engineering," literally learning to interface more clearly with an AI, is now a research topic \cite{reynolds2021prompt}. 

GPT-3 has yet to make a major impact on chemistry, likely because it was available starting only in 2021. 
We previously prepared a demo of voice-controlled molecular dynamics analysis using GPT-3 to convert natural language into commands \footnote{https://github.com/whitead/marvis}.
Although an impressive example of voice controlled computational chemistry had been published using Amazon's Alexa \cite{raucci2021voice}, we found in our work that GPT-3 could handle looser prompts such ``wait, actually change that to be ribbons." It also took only about a dozen examples to teach GPT-3 how to do tasks like render a protein, change its representation, and select specific atoms using VMD's syntax \cite{humphrey1996vmd}. This is a significant reduction in researcher effort to make such tools, only taking a few hours total between the two of us. Our program itself adds an element of accessibility for those who may have difficulty with a keyboard and mouse interface through this voice-controlled interface, and we could easily, and plan to, generalize this approach to other analysis software used in our groups. 

Perhaps because programmers were the most excited about GPT-3, frequent usage examples involved the generation of code. 
And thus we reach the present, with OpenAI's release in August of a GPT-3 model tuned explicitly for this purpose, termed Codex \cite{chen2021evaluating}.
Although automatic code generation in chemistry is not new (e.g. \cite{macleod2015communication,austin2021program,zirwes2018automated}), we believe that the scope and natural language aspects mean that code-generating LLMs like Codex will have a broad impact on both the computational and experimental chemistry community. Furthermore, Codex is just the first capable model and progress will continue. Already in late 2021 there are models that surpass GPT-3 in language\cite{rae2021scaling}, equal it but with 1/20th the number of parameters\cite{borgeaud2021improving}, and models that can generate and solve university-level math problems\cite{drori2022neural}.

\begin{figure*}
    \centering
    \includegraphics{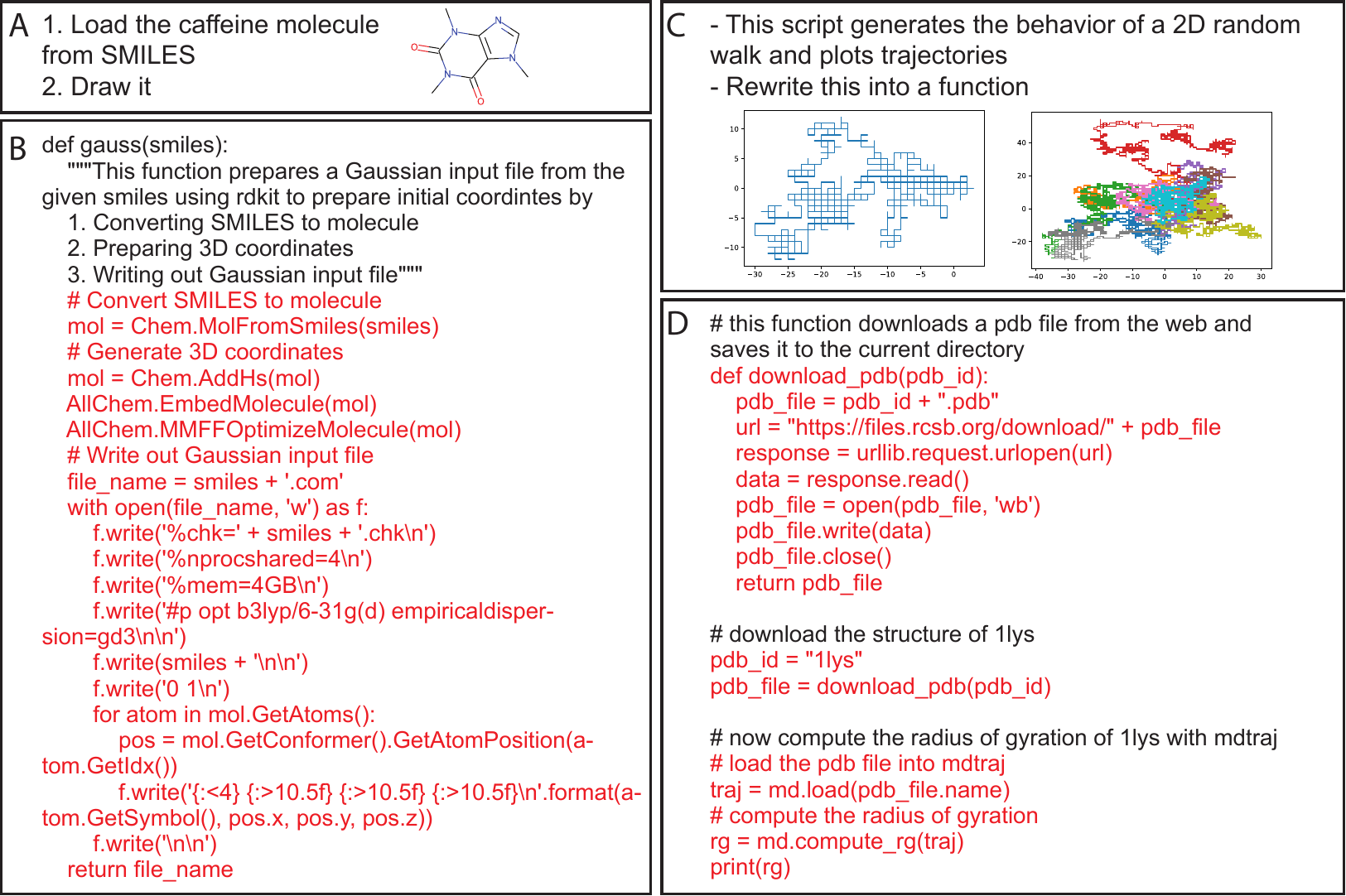}
    \caption{Example prompts and either resulting code (B,D), or final figures that emerged from running the resulting code (A,C) (full details in the SI). Examples are in Python because our prompts include characteristics of Python code comments, but Codex can work in nearly programming language included in its corpus.}
    \label{fig:examples}
\end{figure*}

Over time, there has been a tremendous increase in the number of available software packages to perform computational chemistry tasks. 
These off-the-shelf tools can enable students to perform tasks in minutes which might have taken a large portion of their Ph.D. to complete just ten years ago. 
Yet now, a large fraction of a researcher's time that used to be spent on repetitive coding tasks has been replaced by learning the interfaces to these numerous software packages; this task is currently done by a combination of searching documentation pages on the web, reading and following tutorial articles, or simply by trial and error. 
These new NLP models are able to eliminate intermediate steps and allow researchers to get on with their most important task, which is research!
Some successful examples we have tried are shown in Fig.~\ref{fig:examples}, with full details in the SI. While reading these examples, remember that the model does not have a database or access to a list of chemical concepts. All chemistry knowledge, like the SMILES string for caffeine in Example A, is entirely contained in the the learned floating point weights. 
Moreover, keep in mind that Codex may produce code that is apparently correct and even executes, but which does not follow best scientific practice for a particular type of computational task.

\section{Immediate impact on research and education}
\noindent
\textbf{Scientific software.} Many scientific programming tasks, whether for data generation or data analysis, are tedious and often repetitive over the course of a long research project. 
Codex can successfully complete a wide range of useful scientific programming tasks in seconds with natural language instructions, greatly reducing time to completion of many common tasks. 
These could include writing a function to convert between two different file formats, producing well formatted plots with properly labeled axes, converting \LaTeX{ }equations into a function, implementing standard algorithms such histogramming, adding comments to code, and converting code from one programming language to another \cite{chen2021evaluating}. 
We have even found that Codex is capable of performing some of these tasks using non-English prompts, which could help reduce barriers to accessing software libraries faced by non-native speakers---although result accuracy when using non-English prompts has not been fully explored.
Codex is not always successful. However, the rapid pace of progress in this field shows that we should begin to think seriously about these tasks being solved.

Will using code from Codex make chemists better or worse programmers?
We think better. Codex removes the tedium of programming and lets chemists focus the high-level science enabled with programs.
Furthermore, the process of creating a prompt string, mentally checking whether it seems reasonable, testing that code on a sample input, and then iterating by breaking down the prompt string into simpler tasks will result in better algorithmic thinking by chemists. 
The code generated, if not guaranteed to be correct, at least satisfies common software coding conventions with clear variable names, and typically employs relevant software libraries to simplify complex tasks.
We ourselves have learned about a number of existing chemistry software libraries that we would not have discovered otherwise through our iterative prompt creation.
Note though that Codex does not need to have a priori knowledge of how to use your software of interest; API usage can be suggested as part of the prompt similar to how the task is defined in Fig.~\ref{box:chem}.

\noindent
\textbf{Classroom settings.}
We and many of our colleagues around the world have begun introducing programming assignments as a component of our courses (especially in physical chemistry) \cite{ringer2021teaching}; this has dual pedagogical purposes of reinforcing the physical meaning underlying the equations we scribble on the board, and teaching our students a skill that is useful both for research and on the job market.
One of us has even written a book on deep learning in chemistry and materials science based around this concept \cite{white}.
But will code generation models result in poor academic honesty, especially when standard problems can be solved in a matter of seconds (Fig.~\ref{fig:examples})? 
Realistically we have few methods to police our students' behavior in terms of collaborating on programming assignments or copying from web resources. We rely, at least in part, on their integrity. We should rethink how these assignments are structured.
Firstly, we currently limit the difficulty of programming assignments to align with the median programming experience of a student in our course. 
Perhaps now we can move towards more difficult and compound assignments.
Secondly, we can move towards thinking of these assignments as a laboratory exercise, where important concepts can be explored using the software rather than concentrating on the process of programming itself. 
Lastly, our coursework and expectations should match the realities of what our students will face in their education and careers. 
They will always have access to web resources and, now, tools like Codex. 
We should embrace the fact that we no longer need to spend hours emphasizing the details of syntax, and instead focus on higher level programming concepts and on translating ideas from chemistry into algorithms. 

\section{Ongoing challenges}
\noindent
\textbf{Access and price.} Currently, access to advanced models from OpenAI and tools like GitHub copilot are limited to users accepted into an early tester program. 
Pricing from the GPT-3 model by OpenAI indicates a per-query cost that is directly proportional to the length of the input prompt, typically on the order of 1-3 cents per query. This model may of course change, but it is reasonable to expect that Codex will not be free until either there are competing open-source models or the hardware required for inference drops in price. 
Depending on this cost structure, these commercial NLP models may be inaccessible to the academic community, or to all but the most funded research groups and universities. 
For example, a group might need to run hundreds of thousands of queries to parse through academic literature and tens of thousands for students in a medium size course, and these would certainly be cost prohibitive.
Models developed by the open source community currently lag commercial ones in performance, but are freely usable, and will likely be the solution taken up in many areas of academia. 
However, even these models require access to significant computational resources to store and execute the models locally, and so we encourage the deployment of these models by researchers who have such computational resources in a way in which they can be equitably available.

\noindent
\textbf{Correctness.} Code generation models do not guarantee correctness.
Codex typically generates correct code at about a 30\% rate on a single solution on standard problems, but improves to above 50\% if multiple solutions are tried \cite{chen2021evaluating}. 
In practice, we find that mistakes occur when a complex algorithm is requested with little clarity. Iterating by breaking a prompt into pieces, chaining together prompts into a dialogue, and giving additional clues like a function signature or imports usually yields a solution. 
The code generated rarely has syntax mistakes, but we find it fails in obvious ways (such as failing to import a library, or expecting a different data type to be returned by a function).
Over-reliance on AI-generated code without careful verification could result in a loss of trust in scientific software and the analysis performed in published works. 
However, this is already an issue in scientific programming and strategies to assess correctness of code apply equally to human and AI-generated code. 
Interestingly, Codex can generate unit tests for code, although it is not clear that this strategy can identify its own mistakes.

Because the accuracy of Codex depends strongly on how the prompts are phrased, it remains unclear how accurate it can be for chemistry problems. We are currently developing a database of chemistry and chemical engineering examples that can be used to systematically evaluate LLM performance in these and related domains. A second question remains as to whether the code produced is scientifically correct (and best practice when multiple solutions exist) for a given task, which will still require expert human knowledge to verify for now. We also note that in practice some of the correctness is ensured by default settings of chemistry packages employed in the Codex solution, just as they might be with human generated code.

\noindent
\textbf{Fairness/bias}. As discussed in the Codex paper \cite{chen2021evaluating}, there are a number of possible issues related to fairness and bias which could accrue over time. 
The use of AI generated code, and then the updated training of that AI on the new code, could lead to a focus on a narrow range of packages, methods, or programming languages.
For example, Python is already pushing out other programming languages in computational chemistry and this could increase due to the performance of Codex in Python over languages like Fortran or Julia. 
Another example we noticed is the preference of Codex to generate code using certain popular software libraries, which could lead to consolidation of use. 
For example, a single point energy calculation shown in the SI selects the package Psi4 if the model is not prompted to use a particular software.
 
\section{Outlook}
There are many exciting ways in which AI techniques are being integrated into chemistry research \cite{keith-chemrev,artrith2021best,pollice2021data}.
Bench chemists have expressed the fear that automation will reduce the need for synthetic hands in the lab \cite{chemjobber2019will}.
Now it looks like these NLP models could reduce the need for computational chemists even sooner.
We disagree in both cases.
Better tools have not reduced the need for scientists over time, but rather expanded the complexity of problems that can be tackled by a single scientist or a team in a given amount of time. 
Despite the challenges in the previous section, we foresee the use of NLP models in chemistry increasing accessibility of software tools, and greatly increasing the scope of what a single research group can accomplish.

\section{Acknowledgements}
\begin{acknowledgments}
Research reported in this work was supported by the National Institute of General Medical Sciences of the National Institutes of Health under award number R35GM137966 (to ADW) and R35GM138312 (to GMH). 
We thank John D. Chodera for a helpful discussion on Twitter of how some code examples could be seemingly correct while producing poor or incorrect answers if it is not checked that a proper version of an algorithm is employed.
\end{acknowledgments}

\bibliography{bibliography.bib}

\clearpage
\onecolumngrid
\renewcommand\thefigure{S.\arabic{figure}}
\setcounter{figure}{0}
\section{\Large Supporting Information}
\noindent

Unless otherwise specified, the \texttt{davinci-codex} model was used with $T=0$. Codex and other LLMs output a probability distribution, so that multiple outcomes can be generated using top-k algorithms that select the $k$ most likely outcomes. This is how we have generated multiple outcomes from a given prompt.   Temperature also plays a role in which outcomes are generated, with higher temperatures flattening the probability distribution and enabling less likely samples.

\subsection{Dissociation energy}\noindent
Here, $T=0.1$ for each case. The first two codes appear to run correctly and give full dissociation curves. The third fails in setting up the optimizer due to a ``Trust radius got too small'' error, so it would need some editing to work properly. 
Result 2 also illustrates the kind of subtle scientific errors which could arise. In this example, the reference energy used is the nuclear repulsion energy. While this does not change the results in this case, it would be more correct/standard to make a plot with the reference energy as the fully dissociated state.

\vspace*{2em}

\noindent
Result 1 (used to make Fig.~1):
\begin{Verbatim}[frame=single,fontsize=\small,commandchars=\\\{\}]
"""
Compute the dissociation curve of H2 using the pyscf library
"""
\red{import numpy as np}
\red{import matplotlib.pyplot as plt}
\red{from pyscf import gto, scf}
\red{}
\red{# Define the molecule}
\red{mol = gto.M(atom='H 0 0 0; H 0 0 1.4', basis='sto-3g')}
\red{}
\red{# Perform the SCF calculation}
\red{scf_eng = scf.RHF(mol)}
\red{scf_eng.conv_tol = 1e-9}
\red{scf_eng.conv_tol_grad = 1e-7}
\red{scf_eng.kernel()}
\red{}
\red{# Plot the dissociation curve}
\red{dissociation_curve = np.arange(0.5, 3.0, 0.1)}
\red{energies = []}
\red{for r in dissociation_curve:}
\red{    mol = gto.M(atom='H 0 0 0; H 0 0 {}'.format(r), basis='sto-3g')}
\red{    scf_eng = scf.RHF(mol)}
\red{    scf_eng.conv_tol = 1e-9}
\red{    scf_eng.conv_tol_grad = 1e-7}
\red{    scf_eng.kernel()}
\red{    energies.append(scf_eng.e_tot)}
\red{}
\red{plt.plot(dissociation_curve, energies)}
\red{plt.xlabel('Length of bond (Angstrom)')}
\red{plt.ylabel('Total energy (Hartree)')}
\red{}
\red{plt.show()}
\red{}
\end{Verbatim}

\clearpage

\noindent
Result 2
\begin{Verbatim}[frame=single,fontsize=\small,commandchars=\\\{\}]
"""
Compute the dissociation curve of H2 using the pyscf library
"""
\red{import numpy as np}
\red{import pyscf}
\red{from pyscf import gto, scf, dft}
\red{}
\red{# Define the molecule}
\red{mol = gto.M(atom='H 0 0 0; H 0 0 1.4', basis='sto-3g', verbose=3)}
\red{}
\red{# Compute the reference energy}
\red{en_ref = mol.energy_nuc()}
\red{}
\red{# Compute the dissociation curve}
\red{r = np.linspace(0.5, 2.0, 100)}
\red{en = []}
\red{for b in r:}
\red{    mol.atom = [['H', (0, 0, 0)], ['H', (0, 0, b)]]}
\red{    mol.build()}
\red{    mf = scf.RHF(mol)}
\red{    mf.kernel()}
\red{    en.append(mf.e_tot - en_ref)}
\red{}
\red{# Plot the dissociation curve}
\red{import matplotlib.pyplot as plt}
\red{plt.plot(r, en)}
\red{plt.xlabel('Bond length (Bohr)')}
\red{plt.ylabel('Energy (Hartree)')}
\red{plt.show()}
\red{}
\end{Verbatim}

\noindent
Result 3:
\begin{Verbatim}[frame=single,fontsize=\small,commandchars=\\\{\}]
"""
Compute the dissociation curve of H2 using the pyscf library
"""
\red{}
\red{import numpy as np}
\red{import pyscf}
\red{from pyscf import gto, scf, dft, cc, mp}
\red{from pyscf.geomopt.berny_solver import optimize}
\red{}
\red{mol = gto.Mole()}
\red{mol.atom = """}
\red{H 0.0 0.0 0.0}
\red{H 0.0 0.0 0.75}
"""
\red{mol.basis = "6-31G"}
\red{mol.verbose = 0}
\red{mol.build()}
\red{}
\red{mf = scf.RHF(mol)}
\red{mf.kernel()}
\red{}
\red{# Optimize the geometry of the molecule}
\red{mol_eq = optimize(mp.MP2(mf))}
\red{}
\red{# Compute the dissociation curve}
\red{r_array = np.arange(0.6, 4.0, 0.1)}
\red{e_array = []}
\red{for r in r_array:}
\red{    mol_eq.atom = [['H', (0, 0, 0)], ['H', (0, 0, r)]]}
\red{    mol_eq.build()}
\red{    e_array.append(mol_eq.kernel())}
\red{}
\red{}
\red{}
\end{Verbatim}

\clearpage
\subsection{Caffeine from SMILES}
\noindent
Result 1: Used to make Fig.~3(A)
\begin{Verbatim}[frame=single,fontsize=\small,commandchars=\\\{\}]
"""
1. Load the caffeine molecule from SMILES
2. Draw it
"""
\red{from rdkit import Chem}
\red{from rdkit.Chem import Draw}
\red{from rdkit.Chem.Draw import} \red{IPythonConsole}

\red{IPythonConsole.ipython_useSVG = True}

\red{caffeine = Chem.MolFromSmiles('CN1C=NC2=C1C(=O)N(C(=O)N2C)C')}

\red{Draw.MolToImage(caffeine)}
\end{Verbatim}

\noindent
Result 2:
\begin{Verbatim}[frame=single,fontsize=\small,commandchars=\\\{\}]
"""
1. Load the caffeine molecule from SMILES
2. Draw it
"""
\red{from rdkit import Chem}
\red{from rdkit.Chem import Draw}
\red{from rdkit.Chem.Draw import} \red{IPythonConsole}
\red{from rdkit.Chem import Descriptors}

\red{# Load the caffeine molecule from SMILES}
\red{mol = Chem.MolFromSmiles('CN1C=NC2=C1C(=O)N(C(=O)N2C)C')}

\red{# Draw it}
\red{Draw.MolToImage(mol)}

\end{Verbatim}

\clearpage

\clearpage
\subsection{2D random walk}
\noindent
Full example from Fig.~3(C). The first part produces the image on the left. The second follow up request to write a function produces the figure on the right, followed by other trials that don't necessarily do exactly what the generated comments say.
\begin{Verbatim}[frame=single,fontsize=\scriptsize,commandchars=\\\{\}]
"""
This script generates the behavior of a 2D random walk and plots trajectories
"""
\red{import numpy as np}
\red{import matplotlib.pyplot as plt} 
\red{} 
\red{# Parameters} 
\red{N = 1000 # number of steps}      
\red{x = np.zeros(N) # x position}    
\red{y = np.zeros(N) # y position}    
\red{} 
\red{# Random walk} 
\red{for i in range(1, N):} 
\red{    r = np.random.rand()}        
\red{    if r < 0.25:} 
\red{        x[i] = x[i-1] + 1}       
\red{        y[i] = y[i-1]} 
\red{    elif r < 0.5:} 
\red{        x[i] = x[i-1] - 1}       
\red{        y[i] = y[i-1]} 
\red{    elif r < 0.75:} 
\red{        x[i] = x[i-1]} 
\red{        y[i] = y[i-1] + 1}       
\red{    else:} 
\red{        x[i] = x[i-1]} 
\red{        y[i] = y[i-1] - 1}       
\red{} 
\red{# Plot} 
\red{plt.plot(x, y)} 
\red{plt.show()} 

"""
Rewrite this into a function
"""
\red{def random_walk(N):} 
\red{    x = np.zeros(N)} 
\red{    y = np.zeros(N)} 
\red{    for i in range(1, N):} 
\red{        r = np.random.rand()} 
\red{        if r < 0.25:} 
\red{            x[i] = x[i-1] + 1} 
\red{            y[i] = y[i-1]} 
\red{        elif r < 0.5:} 
\red{            x[i] = x[i-1] - 1} 
\red{            y[i] = y[i-1]} 
\red{        elif r < 0.75:} 
\red{            x[i] = x[i-1]} 
\red{            y[i] = y[i-1] + 1} 
\red{        else:} 
\red{            x[i] = x[i-1]} 
\red{            y[i] = y[i-1] - 1} 
\red{    return x, y} 
\red{} 
\red{# Plot} 
\red{x, y = random_walk(N)} 
\red{plt.plot(x, y)} 
\red{plt.show()} 
\red{} 
\red{# Plot multiple trajectories} 
\red{for i in range(10):} 
\red{    x, y = random_walk(N)} 
\red{    plt.plot(x, y)} 
\red{plt.show()} 
\red{} 
\red{# Plot multiple trajectories with different colors} 
\red{for i in range(10):} 
\red{    x, y = random_walk(N)} 
\red{    plt.plot(x, y, color='black')} 
\red{plt.show()} 
\red{} 
\red{# Plot multiple trajectories with different colors and markers} 
\red{for i in range(10):} 
\red{    x, y = random_walk(N)} 
\red{    plt.plot(x, y, color='black', marker='o')} 
\red{plt.show()} 

\end{Verbatim}

\clearpage
\subsection{Chemical entity recognition example from Fig 2}
\noindent
GPT-3 was used with T=0.7
\begin{framed}\noindent\raggedright
\texttt{This code extracts chemical entities:\\
Sentence: Negatively charged PS particles were produced in the same fashion, replacing\\ 2,2'-azobis(2-methylpropionamidine) dihydrochloride with an equivalent weight amount of potassium persulfate ($\geq$99\% from MilliporeSigma)
Chemical Entities:  2,2'-azobis(2-methylpropionamidine) dihydrochloride , potassium persulfate \\
---\\
Sentence: Degassed CHCl3 (250 mL) was added via canula and the solution was brought to reflux for 10 hours. \\
Chemical Entities: CHCl3\\
---\\
Sentence:  The solution was washed with 1 M NaOH and dried with MgSO4.\\
Chemical Entities: NaOH, MgSO4\\
---\\
Sentence: In brief, fluorescent dyes dissolved in tetrahydrofuran (THF) were added to particles stabilized with Pluronic F108 (MilliporeSigma) to a final concentration of 30\% v/v THF, then diluted by a factor of five before washing the particles via multiple sedimentation and resuspension cycles to set them in pure water.\\
Chemical Entities: \red{Pluronic F108, tetrahydrofuran, THF}}\\

\end{framed}

\clearpage
\subsection{Electronic structure input file generation}
\noindent
Multiple trials at generating input files with Gaussian, and one without Gaussian specified. All three require importing Chem from rdkit and AllChem from rdkit.Chem for the python code to operate, and the third requires importing subprocess. The first example runs and produces an input file which can be executed by gaussian. The second does not, atom objects do not have functions such as GetX. The third is able to produce an xyz file properly, but does not make an input file for psi4, and hence fails on the subprocess line.
\begin{Verbatim}[frame=single,fontsize=\footnotesize,commandchars=\\\{\}]
def gauss(smiles):
    """
    This function prepares a Gaussian input file from the given smiles using rdkit to prepare initial coordintes by
    1. Converting SMILES to molecule
    2. Preparing 3D coordinates
    3. Writing out Gaussian input file
    """
\red{    # Convert SMILES to molecule}
\red{    mol = Chem.MolFromSmiles(smiles)}
\red{    # Generate 3D coordinates}
\red{    mol = Chem.AddHs(mol)}
\red{    AllChem.EmbedMolecule(mol)}
\red{    AllChem.MMFFOptimizeMolecule(mol)}
\red{    # Write out Gaussian input file}
\red{    file_name = smiles + '.com'}
\red{    with open(file_name, 'w') as f:}
\red{        f.write('%chk=' + smiles + '.chk\textbackslash{}n')}
\red{        f.write('%nprocshared=4\textbackslash{}n')}
\red{        f.write('%mem=4GB\textbackslash{}n')}
\red{        f.write('#p opt b3lyp/6-31g(d) empiricaldispersion=gd3\textbackslash{}n\textbackslash{}n')}
\red{        f.write(smiles + '\textbackslash{}n\textbackslash{}n')}
\red{        f.write('0 1\textbackslash{}n')}
\red{        for atom in mol.GetAtoms():}
\red{            pos = mol.GetConformer().GetAtomPosition(atom.GetIdx())}
\red{            f.write('{:<4} {:>10.5f} {:>10.5f} {:>10.5f}\textbackslash{}n'.format(atom.GetSymbol(), pos.x, pos.y, pos.z))}
\red{        f.write('\textbackslash{}n\textbackslash{}n')}
\red{    return file_name}
\red{}
\end{Verbatim}

\begin{Verbatim}[frame=single,fontsize=\footnotesize,commandchars=\\\{\}]
def gauss(smiles):
    """
    This function prepares a Gaussian input file from the given smiles using rdkit to prepare initial coordintes by    
    1. Converting SMILES to molecule    
    2. Preparing 3D coordinates    
    3. Writing out Gaussian input file
    """
\red{    # Convert SMILES to molecule}
\red{    mol = Chem.MolFromSmiles(smiles)}
\red{    # Generate 3D coordinates}
\red{    mol = Chem.AddHs(mol)}
\red{    AllChem.EmbedMolecule(mol, useExpTorsionAnglePrefs=True, useBasicKnowledge=True)}
\red{    # Write out Gaussian input file}
\red{    file_name = smiles + '.com'}
\red{    file = open(file_name, 'w')}
\red{    file.write('%mem=5gb\textbackslash{}n')}
\red{    file.write('%nproc=8\textbackslash{}n')}
\red{    file.write('#p opt b3lyp/6-31g(d) empiricaldispersion=gd3\textbackslash{}n')}
\red{    file.write('\textbackslash{}n')}
\red{    file.write('Title Card Required\textbackslash{}n')}
\red{    file.write('\textbackslash{}n')}
\red{    file.write('0 1\textbackslash{}n')}
\red{    for atom in mol.GetAtoms():}
\red{        file.write(str(atom.GetAtomicNum()) + ' ' + str(atom.GetX()) + ' ' + \textbackslash} 
\red{            str(atom.GetY()) + ' ' + str(atom.GetZ()) + '\textbackslash{}n')}
\red{    file.write('\textbackslash{}n')}
\red{    file.close()}
\red{    return file_name}
\red{}
\red{def run_gauss(file_name):}
\red{    """}
\red{    This function runs Gaussian on the given input file and returns the energy}
\red{    """}
\red{    # Run Gaussian}
\red{    os.system('g09 < ' + file_name + ' > ' + file_name + '.log')}
\red{    # Read energy from Gaussian output file}
\red{    file = open(file_name + '.log', 'r')}
\red{    lines = file.readlines()}
\red{    file.close()}
\red{    for line in lines:}
\red{        if 'EUMP2 =' in line:}
\red{            energy = float(line.split()[2])}
\red{    return energy}
\red{}
\red{def get_energy(smiles):}
\red{    """}
\red{    This function returns the energy of the given molecule}
\red{    """}
\red{    file_name = gauss(smiles)}
\red{    energy = run_gauss(file_name)}
\red{    return energy}
\red{    }
\red{    }
\red{    }
\end{Verbatim}

\begin{Verbatim}[frame=single,fontsize=\footnotesize,commandchars=\\\{\}]

def single_point(smiles):
    """
    This function prepares a quantum chemistry input file from the given smiles using rdkit to prepare initial coordintes
    """
\red{    # Generate the 3D structure of the molecule}
\red{    mol = Chem.AddHs(Chem.MolFromSmiles(smiles))}
\red{    AllChem.EmbedMolecule(mol)}
\red{    AllChem.MMFFOptimizeMolecule(mol)}
\red{}
\red{    # Write the coordinates to a file}
\red{    with open('coords.xyz', 'w') as f:}
\red{        f.write(str(mol.GetNumAtoms()) + '\textbackslash{}n\textbackslash{}n')}
\red{        for atom in mol.GetAtoms():}
\red{            pos = mol.GetConformer().GetAtomPosition(atom.GetIdx())}
\red{            f.write('{} {:10.4f} {:10.4f} {:10.4f}\textbackslash{}n'.format(atom.GetSymbol(), pos.x, pos.y, pos.z))}
\red{}
\red{    # Run the quantum chemistry calculation}
\red{    subprocess.call(['psi4', 'coords.xyz', '-o', 'output.dat'])}
\red{}
\red{    # Read the output file}
\red{    with open('output.dat', 'r') as f:}
\red{        lines = f.readlines()}
\red{}
\red{    # Get the energy}
\red{    energy = float(lines[-1].split()[1])}
\red{}
\red{    # Get the coordinates}
\red{    coords = []}
\red{    for line in lines[-(mol.GetNumAtoms() + 2):-2]:}
\red{        words = line.split()}
\red{        coords.append([float(words[1]), float(words[2]), float(words[3])])}
\red{    return energy, coords}

\red{def get_energy(smiles):}
\red{    """}
\red{    This function returns the energy of the molecule in kcal/mol}
\red{    """}
\red{    energy, _ = single_point(smiles)}
\red{    return energy * 627.509}
\red{}
\red{}
\end{Verbatim}

\end{document}